\documentclass[a4paper,11pt]{article}
\usepackage{jcappub}
\usepackage[utf8]{inputenc}
\usepackage{graphicx,color,amssymb,amsmath,mathtools,cases,bm,float,tocbibind,subfigure,upgreek,longtable,geometry,blindtext,hyperref}
\usepackage[format=hang,font=small,textfont=it]{caption}
\allowdisplaybreaks[2]

\usepackage{pifont}  

\title{\bf{PSpectCosmo}: A Pseudo-Spectral Code for Cosmological Dynamics Spanning Inflation and Reheating}
\author[a,b]{Jie Jiang}

\affiliation[a]{Center for Cosmological Constant Problem, Pusan National University, \\Busan 46241, Republic of Korea}
\affiliation[b]{Department of Physics, Pusan National University, \\Busan 46241, Republic of Korea}
\date{}

\emailAdd{jiejiang@pusan.ac.kr}

\abstract{
\texttt{PSpectCosmo} is a high-performance \texttt{C++} program developed to investigate early-universe cosmological dynamics, with a specific emphasis on the inflationary epoch. Utilizing a Fourier-space pseudo-spectral method, \texttt{PSpectCosmo} enables the precise evolution of interacting scalar fields and gravitational waves, ensuring accurate representation of the power spectrum during inflation. This approach overcomes key limitations of finite difference methods, particularly in maintaining consistency between effective and lattice wave vectors. The code employs the adaptive step size velocity-Verlet algorithm for time integration, offering a balance of numerical stability and high precision. Additionally, \texttt{PSpectCosmo} incorporates a robust mechanism to compute convergent energy density, effectively resolving the issue of divergent energy density at the onset of inflation. These capabilities establish \texttt{PSpectCosmo} as a reliable and versatile tool for probing non-linear cosmological phenomena and inflationary dynamics with exceptional accuracy. The code is publicly available at \href{https://github.com/JieJiang-Cosmology/PSpectCosmo.git}{https://github.com/JieJiang-Cosmology/PSpectCosmo.git}.
}

\begin{document}

\maketitle


\section{Introduction}

Non-linearity in cosmology has emerged as a critical area of research, particularly in understanding complex processes in the early universe. The dynamics following cosmic inflation, referred to as reheating \cite{Kofman:1994rk, Kofman:1997yn, Bassett:2005xm, Allahverdi:2010xz, Amin:2014eta}, involve highly non-linear interactions between fields, making it a fertile ground for numerical studies. Lattice simulations have proven to be an effective approach for investigating these dynamics. Over the years, several open-source lattice codes have been developed to model the reheating process, including \texttt{LATTICEEASY} \cite{Felder:2000hq}, \texttt{CLUSTEREASY} \cite{Felder:2007kat}, \texttt{DEFROST} \cite{Frolov:2008hy}, \texttt{CUDAEASY} \cite{Sainio:2009hm}, \texttt{HLATTICE} \cite{Huang:2011gf}, \texttt{PyCOOL} \cite{Sainio:2012mw}, \texttt{GABE} \cite{Child:2013ria}, $\mathcal{C}$\texttt{osmo}$\mathcal{L}$\texttt{attice} \cite{Figueroa:2020rrl, Figueroa:2021yhd}, and the pseudo-spectral code \texttt{PSpectRe} \cite{Easther:2010qz}. These tools have predominantly focused on studying reheating, as the inflationary period has traditionally been modeled using linear perturbation theory. However, recent developments challenge this convention.

One significant focus in contemporary cosmology is the formation of primordial black holes (PBHs) \cite{Zeldovich:1967lct, Hawking:1971ei, Carr:1974nx}, which has drawn increasing attention due to its implications for dark matter and early universe phenomena. Certain inflationary models predict amplification of small-scale curvature perturbations, potentially reaching non-linear scales. When these amplified perturbations reenter the horizon during the post-inflationary era, their gravitational collapse can form PBHs \cite{Zhou:2020kkf}. Moreover, these amplified small-scale perturbations can induce non-linear effects on other scales, as demonstrated in \cite{Cai:2021yvq}. Such scenarios underscore the necessity of revisiting perturbation evolution during inflation, which may involve non-linearities requiring detailed lattice simulations to capture accurately.

In conventional lattice codes, spatial derivatives are typically computed using finite difference methods. While effective in many applications, this approach introduces discrepancies in inflationary studies. Specifically, the discrete Fourier transform (DFT) of the finite difference operator yields an effective wave vector, $\mathbf{k}_{\rm{eff}}$, that differs from the DFT wave vector, $\mathbf{k}_{\rm{Lat}}$, as shown in \cite{Huang:2011gf}. This mismatch distorts the scalar power spectrum, leading to inaccuracies in modeling inflationary dynamics \cite{Caravano:2022epk}. Substituting $\mathbf{k}_{\rm{Lat}}$ with $\mathbf{k}_{\rm{eff}}$ can resolve this issue for scalar perturbations but complicates the modeling of gravitational waves (GWs). Since the GW source term in Fourier space involves convolution, such substitutions fail to accurately represent mode-mode coupling \cite{Huang:2011gf}. The pseudo-spectral method offers a resolution to these challenges, ensuring $\mathbf{k}_{\rm{eff}} \equiv \mathbf{k}_{\rm{Lat}}$ and providing greater accuracy. This makes it a preferable choice for computing spatial derivatives, especially in inflationary studies.

Another critical challenge arises from the energy density calculation at the onset of inflation. At small scales, the energy density formally diverges due to the contribution of quantum vacuum energy, $E = \langle 0 | H | 0 \rangle = \int \frac{d^3 k}{(2 \pi)^3} \frac{\omega_k}{2} = \frac{1}{4 \pi^2} \int k^3 d k = \infty$. During the initial stages of inflation, lattice modes are sub-horizon and represent quantum fluctuations, whose kinetic and gradient energies should be excluded from the total energy density. However, as these modes cross the horizon, they transition from quantum to classical behavior and must be incorporated into the energy density calculations. To address this, a parameter $R$ is introduced to track horizon exit, ensuring an accurate account of the energy density as inflation progresses.

This work aims to address these methodological challenges by leveraging the pseudo-spectral method and refining energy density calculations to enable precise and consistent modeling of inflationary dynamics. By doing so, it provides a robust foundation for exploring non-linear effects during inflation and their implications for the early universe, including PBHs formation and gravitational wave generation.

\section{Pseudo-Spectral Method}

The finite difference method is a local numerical approach that approximates derivatives at a given point by utilizing data from nearby points. The truncation error in this method scales as $\mathcal{O}(N^{-m})$, where $N$ is the number of grid points, and $m$ depends on the approximation order and the smoothness of the solution. Despite its simplicity, this method requires a higher resolution to achieve high accuracy, particularly for smooth solutions.

In contrast, the pseudo-spectral method is a global numerical approach that approximates derivatives by leveraging all points in the computational domain. This method captures the global structure of the solution and achieves significantly higher accuracy for smooth functions. Assuming the solution is infinitely differentiable, the pseudo-spectral method exhibits exponential convergence, with the error decreasing as $\mathcal{O}(c^N)$, where $0 < c < 1$ \cite{Trefethen:2000}. This rapid convergence makes it an attractive choice for problems involving smooth solutions.

The pseudo-spectral method can accommodate both periodic and non-periodic boundary conditions, making it versatile in various applications. For cosmological simulations, where periodic boundary conditions are common, the DFT offers an efficient mechanism for computing derivatives. The procedure to compute the $\nu$-th order derivative using the Fourier approach is as follows:

\begin{itemize}
	\item \textbf{Fourier Transform}: Compute the DFT of $f(x)$, denoted by $\tilde{f}(k)$.
	\item \textbf{Apply Derivative in Fourier Space}: Define $\tilde{w}(k) = (i k)^\nu \tilde{f}(k)$, where $\tilde{w}(N/2) = 0$ if $\nu$ is odd, to handle the Nyquist frequency appropriately.
	\item \textbf{Inverse Transform}: Perform the inverse DFT (iDFT) on $\tilde{w}(k)$ to obtain the $\nu$-th order derivative of $f(x)$.
\end{itemize}

This method leverages the Fast Fourier Transform (FFT) for efficient computation, achieving a computational complexity of $\mathcal{O}(N \log N)$.

\subsection*{Key Advantages of the Pseudo-Spectral Method}

\begin{itemize}
    \item \textbf{High Accuracy for Smooth Solutions:} Achieves exponential convergence for smooth functions, far outperforming finite difference methods in such cases.

    \item \textbf{Efficiency:} Utilizes FFT for rapid computations, enabling efficient handling of large grid sizes.

    \item \textbf{Elimination of Wave Number Ambiguity:} The effective wave vector $\mathbf{k}_{\rm eff}$ is fully consistent with $\mathbf{k}_{\rm Lat}$ as defined by the DFT, ensuring clarity and accuracy. This consistency is particularly critical for simulations of early-universe physics, such as inflation, as it guarantees the correct computation of the power spectrum. Additionally, the challenge of accurately selecting wave numbers in the evolution of gravitational waves (GWs) is completely resolved.

    \item \textbf{Consistency with Mathematical Identities:} Unlike finite difference methods, the pseudo-spectral method preserves essential mathematical properties such as the \textit{Leibniz} rule for differentiation:

    \begin{equation}
    	( f g ) ^ { \prime } = f ^ { \prime } g + f g ^ { \prime } ~ .
    \end{equation}

    This is particularly advantageous in applications involving gauge fields. For example, when computing derivatives in the presence of a gauge field transformation, $A_\mu \to A_\mu(x) - \partial_\mu \alpha(x)$, the pseudo-spectral method naturally accommodates the necessary compensating terms. In contrast, finite difference methods may introduce errors or inconsistencies in such scenarios \cite{Figueroa:2020rrl}.

    \item \textbf{Minimal Numerical Dispersion:} Reduces numerical dispersion in problems involving wave propagation or oscillatory solutions, leading to more accurate results.
\end{itemize}

\subsection*{Applications and Limitations}

\begin{itemize}
    \item \textbf{Applications:}
    \begin{itemize}
        \item Fluid dynamics, plasma physics, and cosmological simulations.
        \item Problems requiring high accuracy for smooth and periodic solutions.
    \end{itemize}
    \item \textbf{Limitations:}
    \begin{itemize}
        \item Less effective for solutions with discontinuities or steep gradients, where Gibbs phenomena can arise. However, such situations are rare in cosmology.  
        \item Best suited for simple geometries like periodic domains or Cartesian grids. Complex geometries may require alternative or hybrid approaches.
    \end{itemize}
\end{itemize}

By harnessing its global approach and leveraging the computational power of FFT, the pseudo-spectral method stands out as a powerful tool for solving differential equations in numerical simulations.

\section{Lattice simulation during inflation}

\subsection{Equation of motion}

We consider a flat \textit{Friedmann-Lemaître-Robertson-Walker} (FLRW) universe, described by the metric 

\begin{equation}
	d s ^ { 2 } = g _ { \mu \nu } d x ^ { \mu } d x ^ { \nu } = - a ( \eta ) ^ { 2 \alpha } d \eta ^ { 2 } + a ( \eta ) ^ { 2 } \delta _ { i j } d x ^ { i } d x ^ { j } ~ ,
\end{equation}

\noindent where $ a ( \eta ) $ is the scale factor. The parameter $ \alpha $ specifies the time coordinate: $ \alpha = 0 $ corresponds to the \textit{cosmic time} $ t $, while $ \alpha = 1 $ idensifies $ \eta $ the \textit{conformal time}, defined by $ \tau \equiv \int \frac { d t ^ { \prime } } { a \left( t ^ { \prime } \right) } $. 

The dynamics of a scalar field $ \phi $ in this background are governed by its equation of motion:

\begin{equation}
	\phi ^ { \prime \prime } + ( 3 - \alpha ) \mathcal{ H } \phi ^ { \prime } - \frac{ \nabla ^ 2 }{ a ^ { 2 ( 1 - \alpha ) } } \phi + a ^ { 2 \alpha } V _ { , \phi } = 0 ~ ,
\end{equation}

\noindent where $ \mathcal{ H } \equiv a ^ { \prime } / a $ is the Hubble parameter, and $ V _ {, \phi } $ represents the derivative of the scalar potential $ V ( \phi ) $. The Hubble parameter is determined by the Friedmann equation:

\begin{equation}
	\mathcal { H } ^ { 2 } \equiv \left( \frac { a ^ { \prime } } { a } \right) ^ { 2 } = a ^ { 2 \alpha } \frac { \rho } { 3 m _ { p } ^ { 2 } } ~ , \quad \frac { a ^ { \prime \prime } } { a } = \frac { a ^ { 2 \alpha } } { 6 m _ { p } ^ { 2 } } [ ( 2 \alpha - 1 ) \rho - 3 p ] ~ .
\end{equation}

\noindent where $ \rho $ and $ p $ are the total energy density and pressure of the universe, respectively, and $ m _ { p } $ is the reduced Planck mass.

\subsection{Initial condition}

To simulate inflationary dynamics, we impose initial condition on a lattice. According to cosmological perturbation theory, the scalar field $ \phi $ can be decomposed into a homogeneous background and perturbations:

\begin{gather}
	\bar{ \phi } ^ { \prime \prime } + ( 3 - \alpha ) \mathcal { H } \bar{ \phi } ^ { \prime } + a ^ { 2 \alpha } V _ { , \bar{ \phi } } = 0 ~ , \\
	\delta \phi ^ { \prime \prime } _ { k } + ( 3 - \alpha ) \mathcal { H } \delta \phi ^ { \prime } _ { k } + \frac{ k ^ 2 }{ a ^ { 2 ( 1 - \alpha ) } } \delta \phi _ { k } + a ^ { 2 \alpha } V _ { , \phi \phi } \delta \phi _ { k } = 0 ~ .
\end{gather}

To eliminate the Hubble friction term $ ( 3 - \alpha ) \mathcal { H } \delta \phi ^ { \prime } _ { k } $, we redefine the perturbation as $ f \equiv a \delta \phi $. Using conformal time ($ \alpha = 1 $), the equation of motion becomes 

\begin{equation}\label{f_EoM}
	f ^ { \prime \prime } + \left( k ^ 2 + a ^ 2 m _ { \phi \phi } ^ 2 - \frac{ a ^ { \prime \prime } }{ a } \right) f = 0 ~ ,
\end{equation}

\noindent where $ m _ { \phi \phi } ^ 2 \equiv V _ { , \phi \phi } $ is the effective mass of the scalar field. 

\subsection*{Quantization and Initial State}

The field $ f ( \eta , \bf{ x } ) $ and its conjugate momentum $ \pi ( \eta , \bf{ x } ) $ are promoted to quantum operators, $ \hat{ f } ( \eta , \bf{ x } ) $ and $ \hat{ \pi } ( \eta , \bf{ x } ) $ , satisfying the commutation relation 

\begin{equation}
	\left[ \hat { f } ( \eta , \mathbf { x } ) , \hat { \pi } \left( \eta , \mathbf { x } ^ { \prime } \right) \right] = i \hbar \delta _ { \mathrm { D } } \left( \mathbf { x } - \mathbf { x } ^ { \prime } \right) ~ , 
\end{equation}

\noindent where the delta function enforces locality. In Fourier space, this relation becomes 

\begin{equation}
\begin{aligned}
	{ \left[ \hat { f } _ { \mathbf { k } } ( \eta ) , \hat { \pi } _ { \mathbf { k } ^ { \prime } } ( \eta ) \right] } & = \int \mathrm { d } ^ { 3 } x \int \mathrm {~d} ^ { 3 } x ^ { \prime } \underbrace { \left[ \hat { f } ( \eta , \mathbf { x } ) , \hat { \pi } \left( \eta , \mathbf { x } ^ { \prime } \right) \right] } _ { i \hbar \delta _ { \mathrm { D } } \left( \mathbf { x } - \mathbf { x } ^ { \prime } \right) } e ^ { - i \mathbf { k } \cdot \mathbf { x } } e ^ { - i \mathbf { k } ^ { \prime } \cdot \mathbf { x } ^ { \prime } } \\ 
	& = i \hbar \int \mathrm {~d} ^ { 3 } x e ^ { - i \left( \mathbf { k } + \mathbf { k } ^ { \prime } \right) \cdot \mathbf { x } } \\ 
	& = i \hbar ( 2 \pi ) ^ { 3 } \delta _ { \mathrm { D } } \left( \mathbf { k } + \mathbf { k } ^ { \prime } \right)
\end{aligned}
\end{equation}

\noindent ensuring that modes with different wave vectors are independent.

In the Heisenberg picture, the solution for $ \hat { f } _ { \mathbf { k } } ( \eta ) $ is expressed as

\begin{equation}\label{f_operator}
	\hat { f } _ { \mathbf { k } } ( \eta ) = f _ { k } ( \eta ) \hat { a } _ { \mathbf { k } } + f _ { k } ^ { * } ( \eta ) \hat { a } _ { - \mathbf { k } } ^ { \dagger } ~ ,
\end{equation}

\noindent where $ f _ { k } ( \eta ) $ is a classical mode function satisfying Eq. \eqref{f_EoM}, and $ \hat { a } _ { \mathbf { k } } ( \eta _ { i } ) $ are annihilation and creation operators. The initial state is chosen to be the Bunch-Davis vacuum, giving 

\begin{equation}
	f _ { k } ( \eta _ i ) = \frac{ \sqrt{ \hbar } }{ \sqrt{ 2 k } } e ^ { - i k \eta _ i } ~ .
\end{equation}

This sets the initial conditions for the perturbation:

\begin{gather}
	\delta \phi _ { k } ( \eta _ i ) = \frac{ f _ { k } }{ a } = \frac{ \sqrt{ \hbar } }{ a \sqrt{ 2 k } } e ^ { - i k \eta _ i } ~ , \\ 
	\delta \phi _ { k } ^ { \prime } ( \eta _ i ) = \frac{ - i k \sqrt{ \hbar } }{ a \sqrt{ 2 k } } e ^ { - i k \eta _ i } - \frac{ \sqrt{ \hbar } \mathcal{ H } }{ a \sqrt{ 2 k } } e ^ { - i k \eta _ i } ~ . 
\end{gather}

\subsection*{Setting initial conditions on the lattice}

To impose these initial conditions in Fourier space, note that the $ \bf{ k } = 0 $ mode in the Fourier space corresponds to the background field and its velocity:

\begin{gather}
	\tilde{\phi} [0] = \phi _ 0 \cdot N ^ 3 \\
	\tilde{\phi} ^ {\prime} [0] = \phi _0 ^ {\prime} \cdot N ^ 3 
\end{gather}

\noindent where $ N ^ 3 $ is the total number of the lattice points, providing normalization due to the DFT definition.

The Fourier space field $ \tilde{ \phi } [ \bf{ k } ] $ and $ \tilde{\phi} ^ { \prime } [ \bf{ k } ] $ are then converted back to position space using the iDFT. This process initializes the field and its velocity across the lattice, enabling simulation of their evolution. 

\subsection{Energy density}

The energy density of the universe in the lattice is defined as:

\begin{equation}
	\rho = E _ K + E_G + E_V ~ , 
\end{equation}

\noindent where $ E _ K $, $ E _ G $, and $ E _ V $ correspond to the kinetic energy, gradient energy, and potential energy, respectively:

\begin{equation}
	E _ K = \frac{1}{2} \langle \phi ^ {\prime } [\mathbf{n}] ^ 2 \rangle ~ , \quad E _ G =  \frac{1}{2 a ^ 2} \langle (\nabla \phi [\mathbf{n}] ) ^ 2 \rangle ~ , \quad E _ V = \langle V(\phi[\mathbf{n}]) \rangle ~ ,
\end{equation}

\noindent where $ \langle \dots \rangle $ represents the volume average over the lattice, and $ \mathbf{n} $ denotes lattice sites.

\subsection*{Energy computation in Fourier space}
 
By leveraging Parseval's theorem in Discrete Fourier Transform (DFT), which states:

\begin{equation}
	\sum _ { \mathbf{ n } } \phi ^ 2 = \frac{ 1 }{ N ^ 3 } \sum _ { \tilde{ \mathbf{n} } } | \tilde{ \phi } | ^ 2 ~ ,
\end{equation}

\noindent the energy density can also be computed efficiently in Fourier space. Using this, the kinetic energy and gradient energy expressions become:

\begin{gather}
	E _ K = \frac{1}{2}  \langle \phi ^ {\prime } [\mathbf{n}] ^ 2 \rangle = \frac{1}{2 N^6 } \sum _ { \tilde{ \mathbf{n} } } | \tilde{ \phi } ^ {\prime} | ^ 2 ~ , \\ 
	E _ G =  \frac{1}{2 a ^ 2} \langle (\nabla \phi [\mathbf{n}] ) ^ 2 \rangle = \frac{1}{2 a ^ 2 N^6} \sum _ { \tilde{ \mathbf{n} } } | \mathbf{ k } ( \tilde{ \mathbf{n} } ) | ^ 2 | \tilde{ \phi } | ^ 2 ~ . 
\end{gather}

\noindent Separating the $ k = 0 $ component in the expressions: 

\begin{gather}
	E _ K = \frac{1}{2 N^6 } \sum _ { \tilde{ \mathbf{n} } } | \tilde{ \phi } ^ {\prime} | ^ 2 = \frac{1}{2 N^6 } \left( | \tilde{ \phi } ^ { \prime } [ 0 ] | ^ 2 + \sum _ { \tilde{ \mathbf{n} } \neq 0 } | \tilde{ \phi } ^ { \prime } | ^ 2 \right) ~ , \\
	E _ G = \frac{1}{2 a ^ 2 N^6} \sum _ { \tilde{ \mathbf{n} } } | \mathbf{ k } ( \tilde{ \mathbf{n} } ) | ^ 2 | \tilde{ \phi } | ^ 2 = \frac{1}{2 a ^ 2 N^6} \left( 0 + \sum _ { \tilde{ \mathbf{n} } \neq 0 } | \mathbf{ k } ( \tilde{ \mathbf{n} } ) | ^ 2 | \tilde{ \phi } | ^ 2 \right) ~ .
\end{gather}

The $ k = 0 $ component of $ E _ K $ corresponds to the background kinetic energy, $ \frac{ 1 }{ 2 } { \phi _ 0 ^ { \prime } } ^ 2 $ , while the gradient energy for the background vanished. This provides a good approximation of the energy density, ensuring the correct evolution of the scale factor as governed by the Friedmann equation. The remaining terms represent the energy density of the quantum perturbation. As the scale decreases (corresponding to larger $ k $), the energy density contributions from the quantum vacuum perturbations diverge. 

However, removing this divergence directly is not feasible, as perturbations that cross the horizon evolve into classical fluctuations, which are critical for accurately simulating the post-inflationary reheating stage.
To distinguish quantum and classical perturbations, we follow the method outlined in \cite{Baumann:2022mni}, introducing a ratio:

\begin{equation}
	R \equiv \frac { \Delta \hat { f } \Delta \hat { \pi } } { | \hat { f } | | \hat { \pi } | } = \frac { \hbar / 2 } { | \hat { f } | | \hat { \pi } | } 
\end{equation}

Here, $ R $ quantifies the relative quantum uncertainty of the perturbation.
\begin{itemize}
	\item Quantum regime: $ R \sim 1 $, indicating significant quantum uncertainty $ k / a H \gg 1 $.
	\item Classical regime: $ ( R \ll 1 ) $, signifying negligible quantum uncertainty as the perturbation exits the horizon $ k / a H \sim 1 $. 
\end{itemize}

\begin{figure}[h!]
    \centering
    \includegraphics[width=0.8\textwidth]{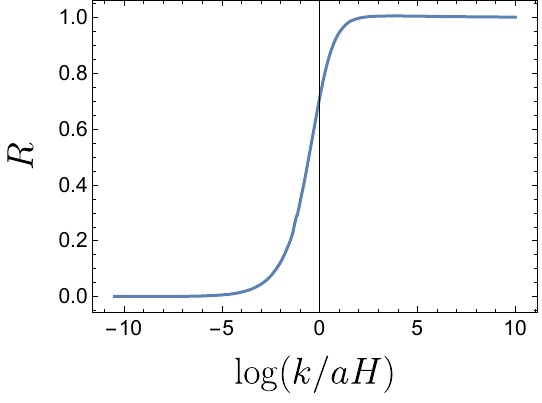} 
    \caption{Evolution of the quantum-to-classical transition ratio $ R $.}
    \label{fig:R}
\end{figure}

The evolution of $ R $ is shown in Fig.~\ref{fig:R}, where the transition from quantum to classical behavior is evident.

While the equations of motion for the perturbation mode functions are classical, the ratio $ R $ provides a clear criterion to identify whether a perturbation behaves quantum mechanically or classically. This distinction is critical for ensuring the accuracy of energy density computations and the proper simulation of post-inflationary dynamics, such as reheating.

\section{Program variable and output}

We use natural units $ c = \hbar = m _ { p } = 1 $ in the code.

In the lattice simulation framework, dimensionless field and space-time variables are utilized as program variables to streamline computations. This redefinition not only enhances numerical stability but also ensures consistency across different scales. The implementation in this code adopts the same strategy outlined in $\mathcal{C}$\texttt{osmo}$\mathcal{L}$\texttt{attice}, as described in \cite{Figueroa:2020rrl}. This approach ensures compatibility with established methodologies and provides a robust foundation for simulation accuracy.

The output files generated by the program are formatted identically to those produced by $\mathcal{C}$\texttt{osmo}$\mathcal{L}$\texttt{attice}, following the conventions detailed in \cite{Figueroa:2021yhd}. This consistency in data structure allows for seamless integration and comparative analysis between simulations, facilitating validation and further development of the code.

Moreover, the computed power spectrum follows the Type I, Version 1 format introduced in \cite{Figueroa:2022}. This standardized representation ensures clarity and compatibility in the interpretation and analysis of spectral data, thereby improving the reproducibility and reliability of the results.

\section{Results}

\begin{figure}[h!]
    \centering
    \includegraphics[width=0.8\textwidth]{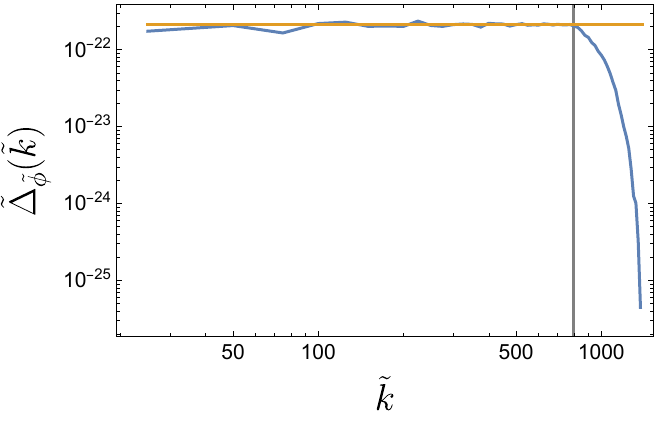} 
    \caption{Dimensionless power spectrum  of $\phi ^ 2 $ model when the perturbation exist horizon. Blue line is the output, orange line is the prediction. The gray is the Nyquist frequency.}
    \label{fig:powerspectrum}
\end{figure}

Figure \ref{fig:powerspectrum} illustrates the dimensionless program power spectrum, $ \tilde{ \Delta } _ { \tilde{ \phi } } ( \tilde{ k } ) $, where $ \tilde{ k } $ represents the dimensionless program wave number in the simulation, for the $ \phi ^ 2 $ model after all perturbation modes have exited the horizon. The blue curve represents the numerical results from the simulation, while the orange curve corresponds to the theoretical prediction. The power spectrum is observed to be nearly scale-invariant, demonstrating excellent agreement with the theoretical expectations.

The vertical gray line marks the Nyquist frequency, the upper limit of resolvable frequencies in the discretized Fourier space. Beyond this frequency, the numerical representation of the power spectrum becomes unreliable because the Fourier components outside the Nyquist range are aliased, resulting in a loss of accuracy. Nevertheless, within the range below the Nyquist frequency, the numerical simulation closely matches the theoretical prediction. This result highlights the precision of the pseudo-spectral method in simulating the inflationary dynamics.

The pseudo-spectral method proves to be an excellent tool for addressing spatial derivatives in cosmological simulations, particularly in the context of inflationary models. Its ability to accurately compute the power spectrum and maintain consistency with theoretical predictions underscores its robustness and effectiveness. This makes the pseudo-spectral method a powerful approach for studying the early universe, providing reliable results that capture the essential physics of inflation with high fidelity.

\section{Summary and outlook}

This code employs the pseudo-spectral method to address challenges in computing the power spectrum that often arise when using finite difference methods for spatial derivatives. By adopting the spectral approach, the code achieves exceptional accuracy while effectively eliminating numerical artifacts inherent to finite difference methods. This capability is particularly critical for cosmological simulations, as demonstrated by the near-perfect agreement between the simulated power spectrum and theoretical predictions. The results underscore the robustness of the pseudo-spectral method, which is ideally suited for capturing the fine-scale dynamics of inflationary cosmology.

In addition to its computational accuracy, the code introduces the diagnostic quantity $ R $, which distinguishes quantum perturbations from classical ones. This distinction ensures that only the classical energy density is considered in the analysis, which is vital for accurately simulating processes such as the post-inflationary reheating stage. The ability to isolate and account for classical perturbations strengthens the code’s utility in studying the evolution of the universe following inflation.

Looking ahead, several key enhancements are planned to expand the code's capabilities and optimize its performance:

1. Transition to GPU-Based Implementation: Moving the code to a GPU platform is a priority for future development. This transition will drastically improve computational efficiency, allowing simulations to handle larger lattice sizes and more intricate physical models in shorter time frames. Such scalability will open doors to higher-resolution studies of inflationary and post-inflationary dynamics.

2. Incorporation of Additional Fields: Expanding the code to include other fields, such as gauge fields, represents a natural progression. The spectral method’s intrinsic adherence to the Leibniz rule makes it particularly adept at handling gauge field interactions. Incorporating such fields will enable the code to investigate a wider array of physical phenomena, including those involving non-Abelian gauge theories, thereby broadening its applicability to complex scenarios in early-universe cosmology.

These planned improvements will further solidify the code as a versatile and precise tool for addressing critical questions in cosmology. By combining cutting-edge numerical methods with computational efficiency and adaptability, this framework has the potential to advance our understanding of the fundamental processes shaping the early universe.

\acknowledgments

I would like to express my gratitude to Richard Easther, Jinn-Ouk Gong, Misao Sasaki, Yi Wang, Kuan Xu, Dong-han Yeom, and Zihan Zhou for their valuable discussions. I extend special thanks to the $\mathcal{C}$\texttt{osmo}$\mathcal{L}$\texttt{attice} school organized by the $\mathcal{C}$\texttt{osmo}$\mathcal{L}$\texttt{attice} team. This work was supported by the National Research Foundation of Korea (Grant No. 2021R1C1C1008622).

\bibliographystyle{JHEP}

\bibliography{ref.bib}{}


\end{document}